# Weyl anomaly for Wilson surfaces

Måns Henningson[†1] and Kostas Skenderis[⋆2]

[†] *Institute of Theoretical Physics, Chalmers University of Technology,
S-412 96 Göteborg, Sweden*

[⋆] *Spinoza Institute, University of Utrecht,
Leuvenlaan 4, 3584 CE Utrecht, The Netherlands*

**Abstract:** We consider a free two-form in six dimensions and calculate the conformal anomaly associated with a Wilson surface observable.

## 1 Introduction

One of the most fascinating results obtained during the course of investigating string dualities is the existence of new interacting theories in six dimensions. These theories possess no dynamical gravity; nevertheless they exhibit many properties of string theory. Such theories are obtained by considering a number of coincident fivebranes and taking the limit of decoupling the bulk gravity.

An example of such a theory is the $(2,0)$ superconformal theory obtained in the decoupling limit of M-theory fivebranes. For several coincident branes, the theory is an interacting conformal field theory about which rather little is known. However, in the large $N$ limit there is a gravitational description through the $AdS/CFT$ correspondence, and one finds that the theory contains $N^3$ degrees of freedom. (This can be obtained either by entropy considerations [1][2] or by calculating the conformal anomaly of the partition function[3]). It follows that the non-abelian system cannot have a conventional gauge theory description since $N^3 \gg \dim GL(N)$. The conformal anomaly of observables associated with closed two-dimensional submanifolds have also been calculated [4] in the large $N$ limit.

By contrast, the world-volume theory on a single $M$-theory fivebrane is well understood and consists of a free $(2,0)$ tensor multiplet, i.e. a two-form gauge field with self-dual field strength, five scalars, and eight spinors. Because of the self-duality constraint on the two-form, there is no covariant action for this field, but the theory can still be defined in a manner similar to how chiral bosons in two dimensions are treated [5]. For chiral bosons, we of course know how to write down an interacting theory, namely the Wess-Zumino-Witten model. So one possible avenue towards understanding the interacting $(2,0)$ theory might be to make a similar generalization of the chiral 2-form theory.

In this letter, we will consider the theory of a free non-chiral two-form $B_{\mu\nu}$ on a six dimensional manifold with coordinates $X^\mu$, $\mu = 1, \ldots, 6$. The classical action is

$$S_0 = -\frac{1}{12} \int d^6 x \sqrt{G} H_{\lambda\mu\nu} H_{\rho\sigma\tau} G^{\lambda\rho} G^{\mu\sigma} G^{\nu\tau}, \tag{1}$$

where $H_{\lambda\mu\nu} = 3D_{[\lambda} B_{\mu\nu]} = \partial_\lambda B_{\mu\nu} + \partial_\mu B_{\nu\lambda} + \partial_\nu B_{\lambda\mu}$ and $D$ is the covariant derivative compatible with the background metric $G_{\mu\nu}$. The field strength $H_{\lambda\mu\nu}$, and thus the action, is invariant under gauge transformations acting as $B_{\mu\nu} \to B_{\mu\nu} + \Delta B_{\mu\nu}$, where $\Delta B_{\mu\nu} dx^\mu \wedge dx^\nu$ is a closed two-form with integer periods. Given a closed two-dimensional embedded submanifold $\Sigma$ with coordinates $\sigma^\alpha$, $\alpha = 1, 2$, we can construct a gauge-invariant Wilson surface observable as

$$W(\Sigma) = \exp 2\pi i \int_\Sigma d\sigma^\alpha \wedge d\sigma^\beta \partial_\alpha X^\mu \partial_\beta X^\nu B_{\mu\nu}. \tag{2}$$

Because of the gauge invariance, we need to introduce a ghost $c_\mu$, an antighost $b^\mu$, and a ghost-for-ghost $\eta$. The latter is necessary since the gauge algebra is reducible. The precise form of the ghost action can be easily obtained using the antifield formalism. However, since the theory is free the ghosts decouple and the precise form of the ghost action is not necessary. To gauge fix we use the covariant gauge fixing condition, $D^\mu B_{\mu\nu} = 0$ and add a gauge fixing term so that the action becomes

$$S = \int d^6 x \sqrt{G} [-\frac{1}{12} H_{\lambda\mu\nu} H_{\rho\sigma\tau} G^{\lambda\rho} G^{\mu\sigma} G^{\nu\tau} - \frac{\alpha}{2}(D^\mu B_{\mu\nu})^2] \tag{3}$$

with some parameter $\alpha$. A change of $\alpha$ will not affect the correlation functions of gauge invariant observables, and henceforth we will take $\alpha = 1$.

Both the classical action (1) and the observable (2) are invariant under conformal transformations acting as

$$\begin{aligned} \delta G_{\mu\nu} &= 2\phi G_{\mu\nu} \\ \delta B_{\mu\nu} &= 0, \end{aligned} \qquad (4)$$

where $\phi$ is an arbitrary infinitesimal parameter function. This invariance is broken by the gauge-fixing term, but the expectation values of Wilson surface observables are still formally invariant. Indeed, we have $\langle W(\Sigma) \rangle = \exp\left(-4\pi^2 I\right)$, where

$$\begin{aligned} I &= \int_\Sigma d\sigma^\alpha \wedge d\sigma^\beta \partial_\alpha X^\rho(\sigma) \partial_\beta X^\sigma(\sigma) \int_\Sigma d\hat\sigma^\gamma \wedge d\hat\sigma^\delta \hat\partial_\gamma X^\mu(\hat\sigma) \hat\partial_\delta X^\nu(\hat\sigma) \\ &\quad \Delta_{\rho\sigma;\mu\nu}(X(\sigma), X(\hat\sigma)). \end{aligned} \qquad (5)$$

Here $\Delta_{\rho\sigma;\mu\nu}(X, X') = \langle B_{\rho\sigma}(X) B_{\mu\nu}(X') \rangle$ is the propagator for the $B_{\mu\nu}$-field. Since the conformal invariance of the action is only broken by the gauge fixing term, the conformal variation of the propagator must be given by an exact term, i.e.

$$\delta \Delta_{\rho\sigma;\mu\nu}(X, X') = \partial_{[\rho} \Lambda_{\sigma];\mu\nu}(X, X') + \partial'_{[\mu} \Lambda'_{|\rho\sigma|;\nu]}(X, X') \qquad (6)$$

for some $\Lambda_{\sigma;\mu\nu}$ and $\Lambda'_{\rho\sigma;\nu}$, where the prime in the derivative indicates that this is a derivative with respect to the $X'$ variable. Since the integral of an exact form over a closed manifold vanishes, it would seem that $I$ is conformally invariant.

However, in the quantum theory divergences arise and have to be regularized and canceled. This can be done in a covariant way, but conformal invariance is generally lost. The objective of this letter is to carry out this procedure and compute the conformal anomaly of this theory. In the next section, we outline the regularization procedure, in section 3, we discuss the form of possible conformal anomalies, and in the last section, we present a detailed computation of the anomaly.

## 2 The regularization procedure

The integrand in (5) divergences along the diagonal of $\Sigma \times \Sigma$. To regularize it in a covariant manner, we exclude from the integration region points with geodesic distance from the diagonal less than some cut-off distance $\epsilon$, i.e. we consider the regulated quantity

$$\begin{aligned} I_\epsilon &= \int_\Sigma d\sigma^\alpha \wedge d\sigma^\beta \partial_\alpha X^\rho(\sigma) \partial_\beta X^\sigma(\sigma) \int_\Sigma d\hat\sigma^\gamma \wedge d\hat\sigma^\delta \hat\partial_\gamma X^\mu(\hat\sigma) \hat\partial_\delta X^\nu(\hat\sigma) \\ &\quad \Theta\left(s^2(X(\sigma), X(\hat\sigma)) - \epsilon^2\right) \Delta_{\rho\sigma;\mu\nu}(X(\sigma), X(\hat\sigma)), \end{aligned} \qquad (7)$$

where $s^2(X, X')$ is the square of the geodesic distance and $\Theta(t)$ is the step function. On general grounds, we expect that

$$I_\epsilon = \epsilon^{-2} I_2 + \log \epsilon I_0 + I_{\text{finite}} + \mathcal{O}(\epsilon), \qquad (8)$$

where $I_2$ and $I_0$ are given by some local expressions integrated over $\sigma$. The divergences can thus be canceled by local counterterms so that we are left with $I_{\text{finite}}$, which however will not be conformally invariant in general. Indeed, under a conformal transformation we have

$$\delta I_\epsilon = \epsilon^{-2} \mathcal{A}_2 + \mathcal{A}_0 + \mathcal{O}(\epsilon), \qquad (9)$$

where the anomaly $\mathcal{A}_0$ equals the conformal variation of $I_{\text{finite}}$. (The coefficient $I_0$ of the logarithmic divergence of $I_\epsilon$ is always conformally invariant.)

We are thus interested in the conformal variation of $I_\epsilon$. It is given by

$$\begin{aligned}\delta I_\epsilon &= \int_\Sigma d\sigma^\alpha \wedge d\sigma^\beta \partial_\alpha X^\rho(\sigma)\partial_\beta X^\sigma(\sigma) \int_\Sigma d\hat{\sigma}^\gamma \wedge d\hat{\sigma}^\delta \hat{\partial}_\gamma X^\mu(\hat{\sigma})\hat{\partial}_\delta X^\nu(\hat{\sigma}) \\ &\quad [\delta(s^2(X(\sigma), X(\hat{\sigma})) - \epsilon^2)\delta s^2(X(\sigma), X(\hat{\sigma}))\Delta_{\rho\sigma;\mu\nu}(X(\sigma), X(\hat{\sigma})) \\ &\quad + \Theta(s^2(X(\sigma), X(\hat{\sigma})) - \epsilon^2)\delta\Delta_{\rho\sigma;\mu\nu}(X(\sigma), X(\hat{\sigma}))] .\end{aligned} \quad (10)$$

Using (6) and partially integrating the second term, this can be written as

$$\begin{aligned}\delta I_\epsilon &= \int_\Sigma d\sigma^\alpha \wedge d\sigma^\beta \partial_\alpha X^\rho(\sigma)\partial_\beta X^\sigma(\sigma) \int_\Sigma d\hat{\sigma}^\gamma \wedge d\hat{\sigma}^\delta \hat{\partial}_\gamma X^\mu(\hat{\sigma})\hat{\partial}_\delta X^\nu(\hat{\sigma}) \\ &\quad \delta(s^2(X(\sigma), X(\hat{\sigma})) - \epsilon^2) \left[\delta s^2(X(\sigma), X(\hat{\sigma}))\Delta_{\rho\sigma;\mu\nu}(X(\sigma), X(\hat{\sigma})) \right. \\ &\quad \left. - \partial_{[\rho}s^2(X(\sigma), X(\hat{\sigma}))\Lambda_{\sigma];\mu\nu}(X(\sigma), X(\hat{\sigma})) - \partial'_{[\mu}s^2(X(\sigma), X(\hat{\sigma}))\Lambda'_{|\rho\sigma|;\nu]}(X(\sigma), X(\hat{\sigma}))\right] .\end{aligned} \quad (11)$$

## 3  The form of the anomaly

The quadratic divergent part of $\delta I_\epsilon$ can be removed by a covariant counterterm (as we will see). The finite piece is the conformal anomaly. Only special $\mathcal{A}_0$ can be removed by counterterms. Since we are considering infinitesimal transformations, $\mathcal{A}_0$ is linear in the conformal factor $\phi$. Let us consider first the case of a constant $\phi$, i.e. we set to zero the terms proportional to the derivative of $\phi$. The remaining terms are the ones that appear in the logarithmic divergence of the correlation function. They are expected to satisfy a Wess-Zumino consistency condition, i.e. the integral of their conformal variation should vanish. (This may be proven using the method of [6], but we have not carried out such an analysis). This means that the integrand is either a conformal invariant or it transforms into a total derivative. Following [7], we call type A the anomalies that transform into a total derivative but themselves are not total derivates and type B the ones that are conformally invariant. We further call type C (C for counterterm) the ones that are themselves total derivatives. Any counterterms will produce terms of this form [3]. We expect the converse to also be true, i.e. any term which is a total derivative of a covariant expression can be removed by a counterterm. In our case, there are two possible type A anomalies, and one type B anomaly, namely

$$\text{Type A}: \quad R_{(2)}, \quad (\nabla^2 X)^2 - 4g^{\alpha\beta}P_{\alpha\beta}, \quad \text{Type B}: \quad \delta^{\alpha\gamma}\delta^{\beta\delta}W_{\alpha\beta\gamma\delta}, \quad (12)$$

where $P_{\alpha\beta}$ is the pull-back of the tensor $P_{\mu\nu} = \frac{1}{4}(R_{\mu\nu} - \frac{1}{10}R\eta_{\mu\nu})$. Let us now consider the terms that are proportional to the derivative of the conformal factor. If we would integrate over the whole manifold (i.e. consider the usual conformal anomalies), then by partial integration these terms would be converted into a Type C anomalies (i.e. ones that can be removed by counterterms). In our case, however, we only integrate over $\Sigma$. Therefore, the terms that involve derivatives of the conformal factor along the normal directions cannot be converted into Type C anomalies. Let us call these terms type D. The most general form of type D terms is fixed by dimensional analysis and covariance to be

$$\text{Type D}: \quad D_\mu \phi \nabla^2 X^\mu, \quad (13)$$

where $\nabla^2 X^\mu$ is the mean curvature vector of $\Sigma$ (which is equal to the trace of the second fundamental form; we review the geometry of submanifolds below). Recall that the components of the mean curvature vector tangential to $\Sigma$ vanish, so (13) contains only the derivative of $\phi$ in the normal directions. Furthermore, it is also easy to see that there is no possible type C anomaly (the only candidate consistent with covariance and dimensions is (13) with $\mu \to \alpha$, but this vanishes as we have just remarked). We therefore conclude that the most general form of the anomaly $\mathcal{A}_0$ is a combination of (12) and (13). We will see that our anomaly is indeed given by such combination.

## 4 The computation

To calculate (11), we need the propagator and its conformal variation. These can be obtained at once by the following method. We first perform a conformal transformation in the gauge fixed action (3) and then invert the kinetic operator. The terms that do not depend on the conformal factor give the propagator, and the others give its conformal variation. In this way we obtain the finite conformal variation, but we will use only the infinitesimal version. Let us perform the conformal transformation $G_{\mu\nu} \to e^{2\phi} G_{\mu\nu}$. The action (3) becomes

$$S = \int d^6x \sqrt{G} [-\frac{1}{12} H_{\lambda\mu\nu} H_{\rho\sigma\tau} G^{\lambda\rho} G^{\mu\sigma} G^{\nu\tau} - \frac{1}{2}(D^\mu B_{\mu\nu})^2$$
$$+ 2 D_\kappa B_{\mu\nu} B_{\rho\sigma} D_\tau \phi G^{\kappa\mu} G^{\nu\rho} G^{\sigma\tau} - 2 B_{\mu\nu} B_{\rho\sigma} D_\kappa \phi D_\tau \phi G^{\mu\rho} G^{\nu\kappa} G^{\sigma\tau}]. \tag{14}$$

After some manipulations we obtain[3],

$$S = \int d^6x \sqrt{G} [\frac{1}{4} B_{\mu\nu} [G^{\mu\rho} G^{\nu\sigma} (D^\tau D_\tau + \frac{1}{10} R) + R^{\mu\rho} G^{\nu\sigma} - 2 W^{\mu\rho\nu\sigma}] B_{\rho\sigma}$$
$$+ 2 D_\kappa B_{\mu\nu} B_{\rho\sigma} D_\tau \phi G^{\kappa\mu} G^{\nu\rho} G^{\sigma\tau} - 2 B_{\mu\nu} B_{\rho\sigma} D_\kappa \phi D_\tau \phi G^{\mu\rho} G^{\nu\kappa} G^{\sigma\tau}] \tag{15}$$

We are interested in the short distance expansion of the propagator in a background with metric $G_{\mu\nu}$. It is convenient to work with Riemann normal coordinates. In these coordinates the metric has the following expansion,

$$G_{\mu\nu} = \eta_{\mu\nu} + \frac{1}{3} R_{\mu\rho\nu\sigma}(X')(X - X')^\rho (X - X')^\sigma + \mathcal{O}((X - X')^3). \tag{16}$$

Expanding the kinetic operator in Riemann normal coordinates and then inverting it, we obtain the short distance expansion of the propagator and its conformal variation. (In the final expression, indices are lowered and raised by the flat metric $\eta_{\mu\nu}$; in the expression below we raised the last two indices in order to display clearly the various antisymmetrizations.) We get

$$\Delta_{\rho\sigma}{}^{\mu\nu}(X, X') = \frac{-1}{4\pi^3} \frac{1}{|X-X'|^4} \left[ \eta_{[\rho}{}^\mu \eta_{\sigma]}{}^\nu + \frac{4}{3} P_{[\rho}{}^{[\mu} \eta_{\sigma]}{}^{\nu]} |X-X'|^2 \right.$$
$$- \frac{1}{3} (X-X')^\kappa (X-X')^\lambda \left( P^{[\mu}{}_\kappa \eta_{\lambda[\rho} \eta_{\sigma]}{}^{\nu]} + \eta^{[\mu}{}_\kappa P_{\lambda[\rho} \eta_{\sigma]}{}^{\nu]} \right)$$
$$- \frac{1}{2} W_{[\rho}{}^{[\mu}{}_{\sigma]}{}^{\nu]} |X-X'|^2 - \frac{1}{3} (X-X')^\kappa (X-X')^\lambda W^{[\mu}{}_{\kappa\lambda[\rho} \eta^{\nu]}{}_{\sigma]}$$
$$\left. + \mathcal{O}((X-X')^3) \right] \tag{17}$$

and

$$\Delta_{\rho\sigma}{}^{\mu\nu}(e^{2\phi} G; X, X') - \Delta_{\rho\sigma}{}^{\mu\nu}(G; X, X') = \frac{-1}{4\pi^3} \frac{1}{|X-X'|^4} \left[ \left( \eta_{\kappa[\rho} \eta_{\sigma]}{}^{[\mu} \eta^{\nu]\tau} - \eta_\kappa{}^{[\mu} \eta^{\nu]}{}_{[\rho} \eta_{\sigma]}{}^\tau \right) \times \right.$$

---
[3]Our conventions are as follows $R_{ijk}{}^l = \partial_i \Gamma_{jk}{}^l + \Gamma_{ip}{}^l \Gamma_{jk}{}^p - i \leftrightarrow j$ and $R_{ij} = R_{ikj}{}^k$. The Riemann and Weyl tensor are related as $R_{\mu\rho\nu\sigma} = W_{\mu\rho\nu\sigma} + G_{\mu\nu} P_{\rho\sigma} + G_{\rho\sigma} P_{\mu\nu} - G_{\rho\nu} P_{\mu\sigma} - G_{\mu\sigma} P_{\rho\nu}$, where $P_{\mu\nu} = \frac{1}{4}(R_{\mu\nu} - \frac{1}{10} R \eta_{\mu\nu})$.

$$\times \left(2\partial_\tau \phi + \partial_\tau \partial_\lambda \phi (X-X')^\lambda\right)(X-X')^\kappa$$
$$-\left(\eta^\kappa{}_{[\rho}\eta_{\sigma]}{}^{[\mu}\eta^{\nu]\tau}\partial_\kappa\partial_\tau\phi + 2\eta^{[\mu}{}_{[\rho}\partial_{\sigma]}\phi\partial^{\nu]}\phi\right)|X-X'|^2$$
$$+\mathcal{O}((X-X')^3)\Big], \tag{18}$$

where $P_{\mu\nu} = \frac{1}{4}(R_{\mu\nu} - \frac{1}{10}R\eta_{\mu\nu})$, $|X-X'|^2 = \eta_{\mu\nu}(X-X')^\mu(X-X')^\nu$, and all tensors are understood to be at $X'$. For infinitesimal $\phi$, the conformal variation of the propagator is indeed exact (as we have anticipated since the $\phi$-dependent terms originate from the gauge fixing term). The tensors $\Lambda_{\sigma;\mu\nu}$ and $\Lambda'_{\rho\sigma;\nu}$ appearing in (6) are equal to

$$\Lambda_{\sigma;\mu\nu} = \frac{1}{4\pi^3}\frac{1}{|X-X'|^2}\eta_{\sigma[\mu}\left(\partial_{\nu]}\phi + \frac{1}{2}\partial_{\nu]}\partial_\lambda\phi(X-X')^\lambda + \mathcal{O}((X-X')^2)\right), \qquad \Lambda'_{\rho\sigma;\nu} = \Lambda_{\nu;\rho\sigma}. \tag{19}$$

The final piece that we need is the geodesic distance between two points $X$ and $X'$ in Riemann normal coordinates and its conformal variation. This can be obtained by integrating the geodesic equation. The result is

$$s^2(X,X') = |X-X'|^2 + \mathcal{O}((X-X')^5), \tag{20}$$
$$\delta s^2(X,X') = |X-X'|^2\left(2\phi + \partial_\mu\phi(X-X')^\mu + \frac{1}{3}\partial_\mu\partial_\nu\phi(X-X')^\mu(X-X')^\nu + \mathcal{O}((X-X')^3)\right).$$

To calculate the conformal anomaly we will need some standard facts about the geometry of submanifolds that we now recall. The induced metric is equal to

$$g_{\alpha\beta} = \partial_\alpha X^\mu \partial_\beta X^\nu G_{\mu\nu}. \tag{21}$$

The Christoffel symbols of the induced metric are equal to

$$\Gamma_{(2)\alpha\beta}{}^\gamma = G_{\mu\nu}\partial_\alpha\partial_\beta X^\mu \partial_\delta X^\nu g^{\gamma\delta} + \partial_\alpha X^\mu \partial_\beta X^\nu \partial_\delta X^\lambda \Gamma_{\mu\nu}{}^\kappa G_{\kappa\lambda} g^{\gamma\delta}. \tag{22}$$

The curvature of the induced metric is related to the target space curvature through the Gauss-Godazzi equation,

$$R_{(2)\alpha\beta\gamma\delta} = R_{\alpha\beta\gamma\delta} + G_{\mu\nu}(\Omega^\mu_{\alpha\delta}\Omega^\nu_{\beta\gamma} - \Omega^\mu_{\alpha\gamma}\Omega^\nu_{\beta\delta}), \tag{23}$$

where $R_{\alpha\beta\gamma\delta}$ is the pull-back of the target space curvature $R_{\mu\nu\rho\sigma}$ and $\Omega^\mu_{\alpha\beta}$ is the second fundamental form

$$\Omega^\mu_{\alpha\beta} = \partial_\alpha\partial_\beta X^\mu - \Gamma_{(2)\alpha\beta}{}^\gamma \partial_\gamma X^\mu + \Gamma_{\kappa\lambda}{}^\mu \partial_\alpha X^\kappa \partial_\beta X^\lambda. \tag{24}$$

It follows that

$$g^{\alpha\gamma}g^{\beta\delta}\Omega_{\alpha\beta}\cdot\Omega_{\gamma\delta} = (\nabla^2 X)^2 + R_{(2)} - g^{\alpha\gamma}g^{\beta\delta}W_{\alpha\beta\gamma\delta} - 2g^{\alpha\beta}P_{\alpha\beta}, \tag{25}$$

where $W_{\alpha\beta\gamma\delta}$ is the pull-back of the Weyl tensor $W_{\mu\nu\rho\sigma}$. (Notice that $W_{\mu\nu\rho\sigma}$, but not its pull-back, is traceless).

We are now ready to calculate (11). Expanding in $u^\alpha = \hat\sigma^\alpha - \sigma^\alpha$, we get

$$X^\mu(\hat\sigma) = X^\mu(\sigma) + \partial_\alpha X^\mu(\sigma)u^\alpha + \frac{1}{2}\partial_\alpha\partial_\beta X^\mu(\sigma)u^\alpha u^\beta + \frac{1}{6}\partial_\alpha\partial_\beta\partial_\gamma X^\mu(\sigma)u^\alpha u^\beta u^\gamma + \cdots \tag{26}$$

We furthermore change integration variables from $\hat\sigma$ to $u$ in (11). We will use Riemann normal coordinates for the induced metric as well[4]. In particular, the Christoffel symbols at $X(\sigma)$ are set to zero and and

---
[4] One may bring $g_{\alpha\beta}$ to $\delta_{\alpha\beta}$ at the origin by appropriate transformation of $\sigma$ and $u$. This has the effect of producing a factor of $\sqrt{\partial_\alpha X \cdot \partial_\beta X}$ in the measure of $\sigma$.

their derivatives at $X(\sigma)$ are given by

$$\partial_\alpha \Gamma_{(2)\beta\gamma}{}^\delta = \frac{1}{3}(R_{(2)\alpha\beta\gamma}{}^\delta + R_{(2)\alpha\gamma\beta}{}^\delta). \tag{27}$$

The various quantities entering in the calculation have the following expansion

$$|X(\sigma) - X(\hat{\sigma})|^2 = u^2 - \frac{1}{12}\Omega_{\alpha\beta} \cdot \Omega_{\gamma\delta} u^\alpha u^\beta u^\gamma u^\delta + \mathcal{O}(u^5) \tag{28}$$

$$\delta(s^2 - \epsilon^2) = \frac{1}{2u}\left[\delta(u-\epsilon)(1 + \frac{1}{4}u^2 \Omega_{\alpha\beta} \cdot \Omega_{\gamma\delta} \hat{u}^\alpha \hat{u}^\beta \hat{u}^\gamma \hat{u}^\delta) - \frac{1}{4}\delta'(u-\epsilon)u^3 \Omega_{\alpha\beta} \cdot \Omega_{\gamma\delta} \hat{u}^\alpha \hat{u}^\beta \hat{u}^\gamma \hat{u}^\delta + \mathcal{O}(u^3)\right]$$

$$\epsilon^{\alpha\beta}\epsilon^{\gamma\delta}\partial_\alpha X^\rho(\sigma)\partial_\beta X^\sigma(\sigma)\partial_\gamma X^\mu(\hat{\sigma})\partial_\delta X^\nu(\hat{\sigma})\eta_{\rho\mu}\eta_{\sigma\nu} =$$
$$2 - \left(\nabla^2 X \cdot \Omega_{\alpha\beta} + \frac{2}{3}(R_{(2)\alpha\beta} - \delta^{\gamma\delta}R_{\alpha\gamma\beta\delta})\right)u^\alpha u^\beta + \mathcal{O}(u^3),$$

where $u^\alpha = u\hat{u}^\alpha$.

Putting everything together, and performing the (elementary) $u$ integration, we finally obtain

$$\delta\langle W(\Sigma)\rangle = \exp\int_\Sigma d\sigma^1 d\sigma^2 \sqrt{\det \partial_\alpha X \cdot \partial_\beta X} \left[\phi\frac{4}{\epsilon^2} + \phi\left(-\frac{3}{4}((\nabla^2 X)^2 - 4g^{\alpha\beta}P_{\alpha\beta}) - \frac{1}{2}R_{(2)} - \frac{1}{6}g^{\alpha\gamma}g^{\beta\delta}W_{\alpha\beta\gamma\delta}\right) - \frac{5}{6}\nabla^2 X^\mu D_\mu \phi\right]. \tag{29}$$

Notice that the anomaly is indeed a sum of type A, B and D anomalies. One should also note that it is *not* a multiple of the anomaly obtained in the large $N$ limit of $N$-coincident $M$-theory five-branes [4].

We note that the quadratic divergence can indeed be removed by wave-function renormalization. Namely, rather than $W(\Sigma)$, we should consider the renormalized observable

$$W(\Sigma)_R = W(\Sigma) \times \exp\int_\Sigma d\sigma^1 d\sigma^2 \left(-\frac{2}{\epsilon^2}\sqrt{\det g_{\alpha\beta}}\right)$$
$$= \exp\int_\Sigma d\sigma^1 d\sigma^2 \left[2\pi i\epsilon^{\alpha\beta}\partial_\alpha X^\mu \partial_\beta X^\nu B_{\mu\nu} - \frac{2}{\epsilon^2}\sqrt{\det \partial_\alpha X \cdot \partial_\beta X}\right]. \tag{30}$$

(More precisely, one should also subtract the logarithmic divergences, but we have not displayed these.)

Notice the similarity between the renormalized Wilson observable and the worldsheet string action. Pushing this similarity a step further, it is tempting to identify the cut-off $\epsilon$ with the string length of an underlying string theory, and the coefficient of $R_{(2)}$ term in the conformal anomaly with its central change. Notice also that the conformal anomaly of the (large $N$) interacting $(2,0)$ theory [4] does not contain an $R_{(2)}$ term, indicating a cancellation between the various contributions.

## Acknowledgements

We would like to thank each other's institute for hospitality and support during part of this work. MH is supported by the Swedish Natural Science Research Council (NFR), and KS is supported by the Netherlands Organization for Scientific Research (NWO).